\documentclass[12pt]{iopart}
\usepackage{color}
\usepackage{graphics}
\usepackage{soul}
\usepackage{epsfig}
\usepackage{float}
\usepackage[superscript,biblabel]{cite}
\usepackage[colorlinks]{hyperref}\hypersetup{
     colorlinks   = true,
     citecolor    = blue
}
\usepackage{hyperref}
\hypersetup{
    colorlinks=true,
    linkcolor=blue,
    filecolor=magenta,      
    urlcolor=cyan,
}
\begin{document}
\title[]{Signatures of an energetic charge bunch moving in a plasma \footnote{This manuscript has been authored by UT-Battelle, LLC, under contract DE-AC05-00OR22725 with the US Department of Energy (DOE). The US government retains and the publisher, by accepting the article for publication, acknowledges that the US government retains a nonexclusive, paid-up, irrevocable, worldwide license to publish or reproduce the published form of this manuscript, or allow others to do so, for US government purposes. DOE will provide public access to these results of federally sponsored research in accordance with the DOE Public Access Plan (http://energy.gov/downloads/doe-public-access-plan).}}

\author{Vikram Dharodi$^1$, Atul Kumar$^2$ and Abhijit Sen$^1$}

\address{$^1$Institute for Plasma Research, HBNI, Bhat, Gandhinagar - 382428, India}
\address{$^2$Oak Ridge National Laboratory, Oak Ridge, TN 37831, United States of America}
\ead{vikram.ipr@gmail.com}
\vspace{10pt}

\begin{abstract}
A charge bunch moving in a plasma can excite a variety of linear and nonlinear waves in the form of trailing wakes, fore-wake shocks and precursor solitons. These structures can further interact with the background plasma to create secondary effects that can serve as signatures of the passage of the charge bunch. Using particle-in-cell simulations we investigate in detail the dynamics of a plasma system that is being traversed by an energetic charged ion bunch. Using two different shapes of the charge source, namely, an idealized one dimensional line source and a two dimensional thin rectangular source  we examine the differences in the nature of the excited wave structures and their consequent impact on the background plasma. Our simulations reveal interesting features such as the dependence of the precursor speeds on the total charge of the ion bunch, local particle trapping, and energization of the trapped ions in various regions along the traversal path leading to the formation of energetic ion beam-lets.  The collective excitations and the signatures in the ambient plasma could prove useful in practical applications such as in ion beam heating of plasmas. They can also help in analysing  the trajectories of charged objects like space debris orbiting in the ionosphere.
\end{abstract}

\section{Introduction}
\label{intro}

One of the well known phenomenon associated with the passage of an object in a neutral fluid is the excitation of trailing waves behind the object more popularly known as wakes or wake fields. In a plasma, such wake fields in the form of plasma waves can arise when a charged bunch of electrons or ions or an intense laser pulse travels through it. Plasma wake fields have been intensely studied in recent times because of their potential application as a means of accelerating charged particles. An electron or an ion injected in such a structure and moving with the same speed as the wake experiences a nearly DC electric field and thereby gains energy from it \cite{tajima1979laser}. Since a plasma can sustain large electric fields, these wake fields can accelerate particles to very high energies and the wake field based accelerator scheme has made spectacular progress over the past several years \cite{joshi1984ultrahigh,joshi2018plasma,tajima2020wakefield,kurz2021demonstration}. \\

Apart from these wake fields which trail a moving charge bunch, it is also possible to excite nonlinear wave structures in the fore-wake region i.e. ahead of the bunch. Such a phenomenon, well known in hydrodynamics, has not received much attention in plasma physics, except in very recent years \cite{sen2015nonlinear,kumar2016wakes,jaiswal2016experimental,arora2019effect}.  It has been shown that when a charge bunch moves in an unmagnetized plasma at a speed greater than the ion acoustic speed, a train of nonlinear structures in the form of precursor solitons can arise ahead of the moving charge and move away from it at a faster speed \cite{sen2015nonlinear}. The existence of such precursor solitons has also been experimentally confirmed in controlled experiments carried out in a dusty plasma device with the dust fluid made to flow over a charged obstacle at a speed faster than the dust acoustic speed \cite{jaiswal2016experimental}. In principle, such excitations can arise in a magnetized plasma as well in the form of magnetosonic or Alfvenic solitons. A proof of principle evidence of the existence of such electromagnetic precursor solitons was provided in an earlier publication \cite{kumar2020precursor} using particle-in-cell [PIC] simulations. It was shown that a rigid charged particle source, in the form of a nearly one dimensional line source could excite these solitons when its velocity exceeded the magneto-sonic speed. The simulations also confirmed the existence of electrostatic solitons, by turning off the ambient magnetic field and moving the source at a speed faster than the ion acoustic speed. The focus of the earlier study was on delineating the conditions for the excitation of the ion acoustic and magnetosonic precursors in an unmagnetized and magnetized plasma set-up, respectively. The precursors were also identified as solitons from their propagation characteristics.  The topic of the subsequent interaction of these wave structures with the ambient plasma was not addressed nor was the dependence of the wave properties on the nature of the beam source explored. Our present study is devoted to an investigation of these questions. We investigate in greater detail the nature of the excited structures as a function of the source size. This is done by comparing the results of propagating a thin rectangular shaped source with different transverse extents with those obtained for an idealized one dimensional source.  We also look at the subsequent interactions of the excited waves with the background plasma in various regions along the path of the charged beam. These include the wake region, the density depleted region just behind the moving source and the region ahead of the source where precursor solitons or shock structures propagate. For this we carry out a detailed study of the temporal evolution of the particle density and energy density in various regions as well as some individual particle kinetics in special regions. For simplicity we restrict ourselves to excitations created by a charged ion source propagating in an unmagnetized plasma.

Our simulations show interesting differences for the two cases both in the characteristics of the wave structures and their subsequent interaction with the plasma particles. The finite size of the source is seen to introduce an electromagnetic component in the excited structures. The propagation speed of the precursors is found to have a direct dependence on the strength of the charge in the source. We also observe significant self-injection of ions in the ion-depleted region immediately behind the charge source. These ions experience energization and focusing effects leading to the formation of beam like structures. Ions are also seen to be trapped between precursor structures and to get energized due to multiple bounces off these structures. The trajectories of these particles are analysed using particle tagging techniques. The implications and potential applications of these secondary signatures created by the passage of an energetic ion bunch in the plasma are discussed.

The paper is organized as follows. In section \ref{model} we briefly describe the physical model used for our simulations and also give the essential computational details. Our main results are presented in section \ref{results} with separate subsections devoted to energization due to collective excitations arising from an idealized thin line source  (\ref{1d-source}) and  a two dimensional thin rectangular source (\ref{finite_line}). Section \ref{disc} gives a summary of our results and discusses their possible occurrences in nature as well as their potential applications. 

\section{Simulation model and details}
\label{model} 

\begin{figure}
    \center
    \includegraphics[width=1.0\textwidth]{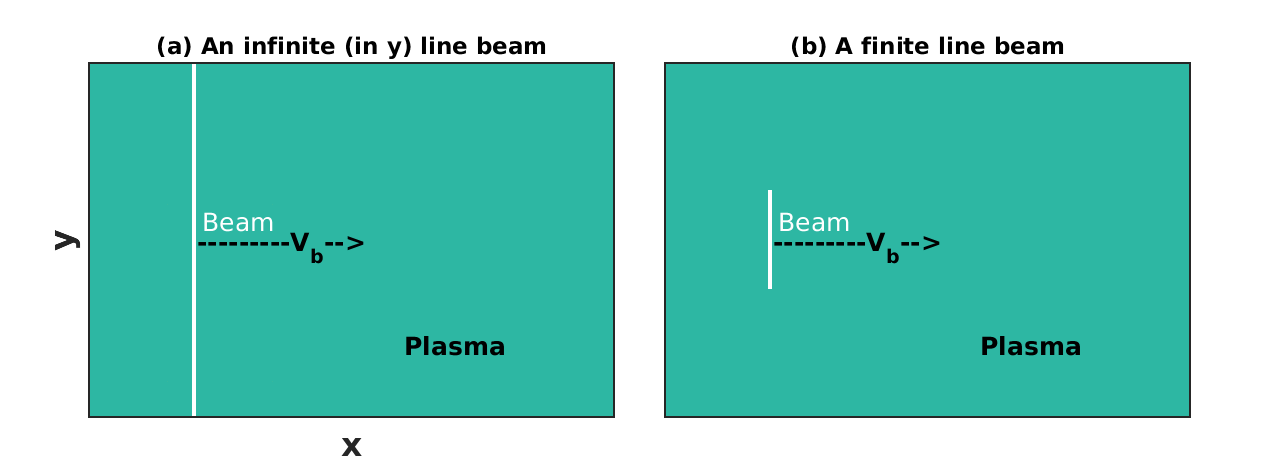} 
    \caption{Schematic diagram (not to scale) of the simulation geometry showing two different shaped charge sources  (a) an infinite (in $y$) thick line source and (b) a finite thin rectangular source.  They move with speed $V_b$ in the positive $\hat{x}$-direction.}  
    \label{fig1}
\end{figure}

\noindent
 We have carried out 2d-3V particle-in-cell simulations using the OSIRIS-4.0 code \cite{hemker2000particle,fonseca2002osiris,fonseca2008one} for a quasi-neutral plasma system where the ions are taken to be cold such that the ion thermal velocity $v_{thi}=8.85{\times}{10^{-5}}c$, while the electrons are warm with a thermal velocity $v_{the}=0.5c$. At this electron temperature the ion acoustic sound speed in the medium is \cite{kumar2020precursor} $V_{cs}=0.1c$  ($V_{cs}=\sqrt{{k_B{T_e}}/{m_i}}={v_{the}/{\sqrt{m_i/m_e}}}$). For simplicity and shorter simulation times the ion to electron mass ratio ${m_i/m_e}$ has been taken to be $25$. We consider a rectangular box of $L_x(=1000{\delta{s}})$~${\times}$~$L_y(=100{\delta{s}})$ in the $ x-y $ plane, where ${\delta{s}}(=c/\omega_{pe})$ is the skin-depth of the plasma. $L_x$ and $L_y$ are the system lengths in the $x$ and $y$ directions. The equilibrium plasma density is taken as $n_0=3.10\times 10^{20} cm^{-3}$. The spatial resolution chosen in the simulation is $20$ cells per electron skin depth with $10$~${\times}$~$10$ particles per cell for each species corresponding to a grid size of $\Delta x = 0.05 c/\omega_{pe}$ and the temporal resolution is given by the time step $ \Delta t = 0.0283 \omega_{pe}^{-1} $. In such a plasma system a short pulsed positive ion charge bunch is made to travel in order to create linear and nonlinear wave excitations in the plasma. We consider two specific shapes of the charge bunch: (i) an infinite (in $y$) and finite thickness (in $x$) line source and (ii) a finite thin rectangular source. Schematic diagrams  corresponding to these two cases are shown in Fig.~\ref{fig1}(a) and Fig.~\ref{fig1}(b), respectively.  
 For all the simulation runs,  the line beam is initially positioned at $x=500{\delta{s}}$  and made to travel at $V_b=0.11c$ that is a bit faster than the ion acoustic speed $V_{cs}=0.1c$. The corresponding Mach number is $M=1.1$ (where $M= V_b/V_{cs}$). For simplicity we have considered a rigid beam that retains its shape and does not change its velocity while propagating through the plasma.
\section{Simulation results and Observations}
\label{results}
The prime objectives of this section are to numerically explore the evolution of wave structures generated using two types of positive ion charge sources shown in figure~\ref{fig1} and their consequent impact on the energizing process. Both types of sources have constant charge density $\rho^{c}_{bm}=1.1$ and the same  width $\delta{x}=1.02$ (along $\hat{x}$ direction). The source length $\delta{y}$ (along $\hat{y}$ direction) is the only parameter that is varied. All the simulations are performed with absorbing boundary conditions in the $\hat{x}$ direction and periodic in the $\hat{y}$ direction for both the particles and fields.

\subsection{Idealized one dimensional thin line source } 
\label{1d-source}
\begin{figure}[t]
\center
\includegraphics[width=1.0\textwidth]{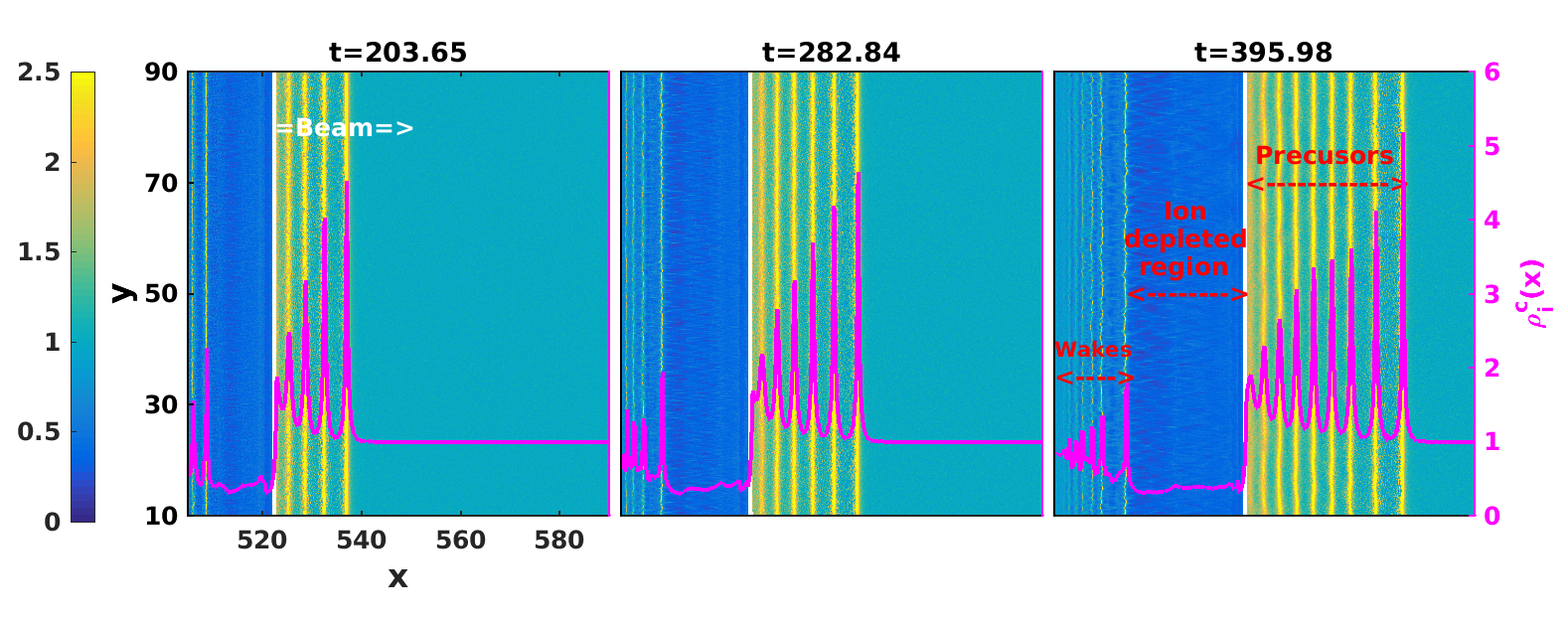}
\caption{Time evolution of the ion charge density $\rho^{c}_{i}(x,y)$ due to passage of an infinite line source beam ($\delta{x}=1.02, \delta{y}=L_y$). The beam creates well developed trailing wake fields and precursor structures moving ahead of it. The line plots superposed on the snapshots are the $y$ averaged density values $\rho^{c}_{i}(x)$ providing a one dimensional profile of the perturbations as a function of $x$ (right y axis).}  
 \label{fig2}
\end{figure}

We first discuss an infinite length beam case, which develops only longitudinal perturbations in the background plasma. Understanding this case allows us to determine the impacts of a finite length beam. For example, it is of interest to know whether the speed of the perturbations is affected by the transverse dimension of the beam and whether it can provide a means of controlling the speed. Figure~\ref{fig2} shows the late-time evolution of the ion charge density $\rho^{c}_{i}(x,y)$ due to the passage of an infinite line source beam (vertical white-solid line) through the plasma.  The solid line (magenta color) over these 2d density snapshots represents its  one dimensional profile $\rho^{c}_{i}(x)$ as a function of $x$ only. $\rho^{c}_{i}(x,y)$ is averaged in the $y$ direction since density is homogeneous in the transverse direction $i.e$
\begin{equation}\label{eqn:}
    \rho^{c}_i(x)={\frac{1}{N_y}}{\int^{L_y}_{0}{{\rho^{c}_{i}(x,y)}dy}}
	\nonumber
\end{equation}  
$N_y$ represents the total number of grid points in the y direction. This travelling beam generates a strong density inhomogeneity in the longitudinal direction.
It is clearly seen that this inhomogeneity is the result of the creation of precursor waves ahead of the beam (yellow-colored disturbances), the formation of a region with a depleted number of ions just behind the beam (a dark-blue-coloured region) and wake structures following this depressed region (cyan-colored disturbances). Such structures were previously observed in the simulations reported by Kumar and Sen \cite{kumar2020precursor} for the case when the ambient magnetic field was taken to be zero. They were obtained to benchmark the code and also to establish connections with previous one dimensional electrostatic results obtained in fluid simulations. In the present paper we examine these structures in greater detail in order to characterize them and to also study their impact on the background plasma.\\

\begin{figure}
\center
\includegraphics[width=1.0\textwidth]{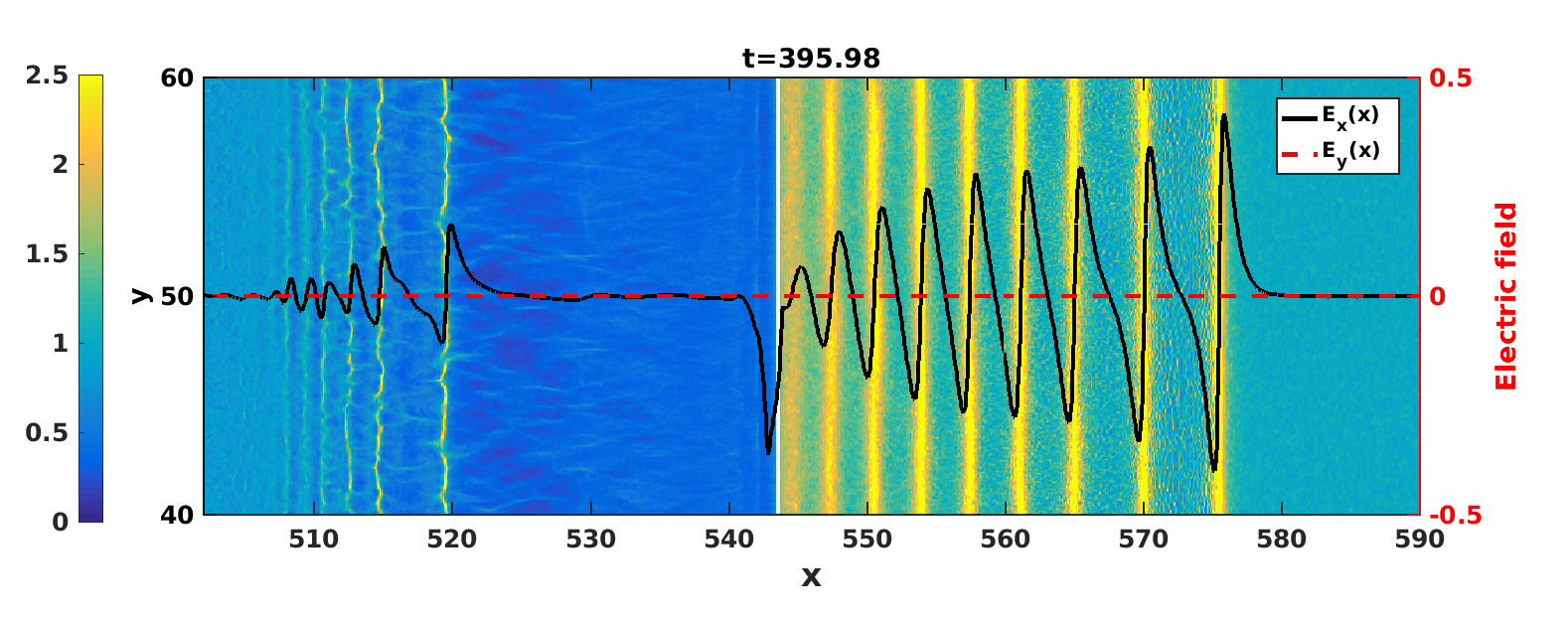}
\caption{Spatial variation of electric field components (longitudinal electric field  $E_x$ and $E_y$) with respect to $x$ at a particular instant of time $t = 395.84$.}
 \label{fig3}
\end{figure}

\begin{figure}[h]
\center
\includegraphics[width=1.0\textwidth]{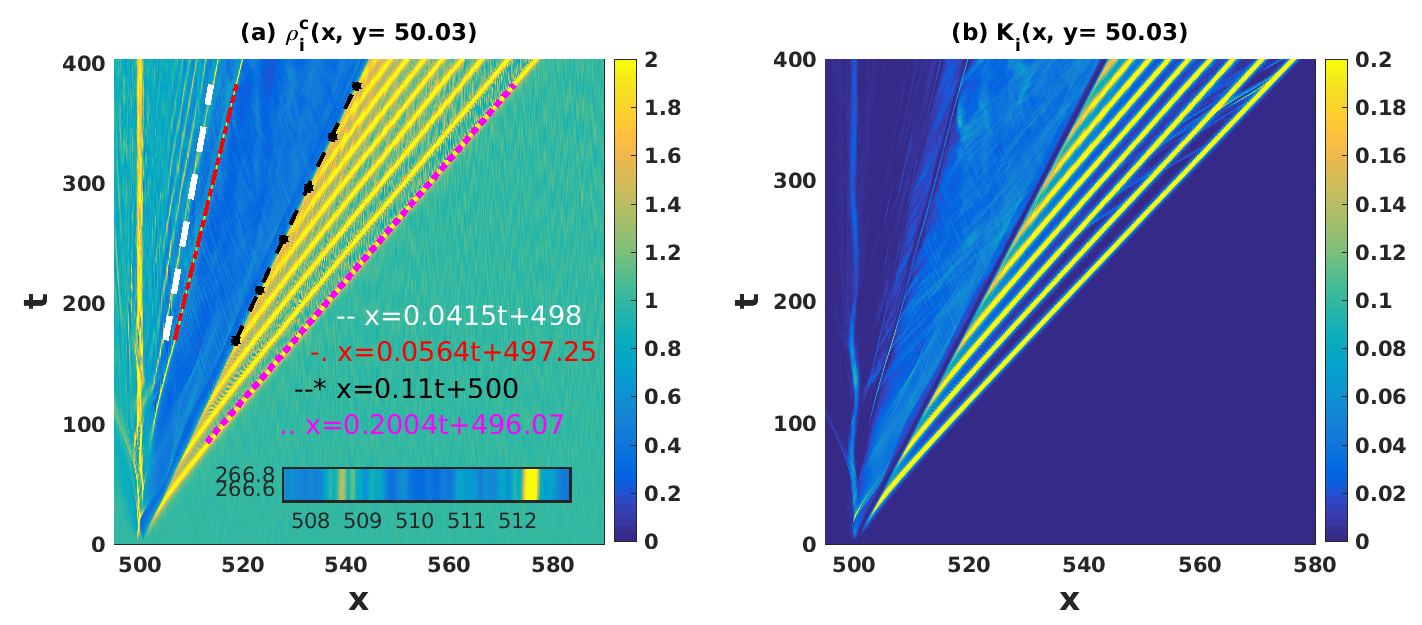}
\caption{The temporal evolution of (a) the ion charge density and (b) the kinetic energy at $y=50.03$}
  \label{fig4}
\end{figure}

To establish the basic character of these structures we first look at the electric field components of these waves. In Fig.~\ref{fig3} we have superposed on the ion charge density colour plot, the longitudinal field component, $E_x(x)$, as a black-solid curve, and the transverse component, $E_y(x)$, as a red-dotted curve, at a time $t=395.98$. For these components of the electric field, the right y-axis defines the scale. As can be seen from this figure, the transverse component is extremely weak compared to the longitudinal component indicating that the waves are primarily electrostatic in nature. Furthermore the density depleted region is nearly free of any electric field and the ion density is uniform in this region. $E_x(x)$ has sharp spikes at either edge of this region that is consistent with the creation of a region of density depression behind the moving source. A careful analysis  of Fig.~\ref{fig3} tells us that the electrostatic  force for ions in the density depleted region is acting in the forward direction and is responsible for the acceleration of the self injected ions in the direction of the beam propagation.
We next look at the time evolution of the various structures described above and the changes in the background plasma as a result of their creation. This is best captured in Fig.~(\ref{fig4}) which shows the temporal evolution of the density and the kinetic energy as a function of $x$ for a fixed value of $y=50.03$. Since in this 1d simulation case there is uniformity in the $y$ direction, this plot provides a comprehensive picture of the spatio-temporal evolution of wave excitations as well as the background density (Fig.~\ref{fig4}(a)) and kinetic energy (Fig.~\ref{fig4}(b)) of the system. From Fig.~\ref{fig4}(a) we can clearly see that with time the beam excites more and more precursor structures ahead of it which move faster than the beam. These precursor structures with enhanced ion densities (yellow color) travel along $x$ at a steady pace with a super sonic velocity of $0.2004 c$. The velocity can be deduced from the straight line fit of $x=0.2004 t + 496.07$ to the trajectory of the precursor. This is consistent with earlier fluid \cite{sen2015nonlinear,kumar2016wakes} and PIC simulations \cite{kumar2020precursor} which have identified these structures as precursor ion acoustic solitons.  Fig.~\ref{fig4}(a) also shows that with the passage of time the region of depressed density keeps on expanding. This can be understood from mass conservation arguments. Basically as the beam generates more and more density enhanced structures (precursor solitons) ahead of it, mass is transferred from the region behind the beam to enable this process and hence the depression region grows in size. Trailing behind the depressed region are the wake structures. Unlike the precursor solitons these are dispersive structures whose amplitudes die down as a function of the distance behind the depressed region. Far away from the depressed region the amplitudes are quite small and the waves travel with the group velocity of the linear ion acoustic wave, namely,
\begin{equation}
    C_{g} = \frac{\partial \omega}{\partial k}= \frac{C_s}{(1 + k^2\lambda_d^2)^{3/2}}
\label{group}
\end{equation}
where $k$ is the wave number and $\lambda_d$ is the electron Debye length. For our simulation parameters, $\lambda_d \sim 0.5 c/\omega_{pi}$. The wave number $k \sim 1.8 \omega_{pi}/c$ so that $k\lambda_d \sim 0.9$. Using (\ref{group}) this yields a group velocity of $\sim 0.04 C_s$. The velocity of these linear wakes can be deduced from a straight line fit to the green lines shown in the far left portion of Fig.~\ref{fig4}(a) and the velocity is found to be $\sim 0.04 c \sim 0.4 C_s$. The lead portion of the wake, seen as a wave structure with a substantial amplitude, that remains attached to the rear end of the depressed region, travels somewhat faster than the linear wakes. A linear fit to its trajectory yields a velocity of $0.0564 c  \sim 0.564C_s  $. The enhanced velocity can be attributed to finite amplitude effects giving the lead wake structure a nonlinear character.  

Finally, in Fig~\ref{fig4}(b), we look at the spatio-temporal evolution of the kinetic energy of the system as the beam passes through it. The energy of the beam and the precursors are predominantly due to their directed velocities with the precursors showing a higher energy content than the beam. There is no significant heating of the background plasma in the fore-wake region. Some traces of weakly energetic ion beam-lets are seen in regions between some of the precursors which could be due to trapping and repeated bouncing of some ions between them. Such an effect becomes more pronounced in 2d simulations which will be discussed in the next section. We also note that the density depressed region shows evidence of background heating of the ions in that region. The weak wake structures do not impact the background energy in any noticeable manner. 

\subsection{A two dimensional thin rectangular source}
\label{finite_line}
We now turn to the case of a two dimensional thin rectangular source
moving through the plasma as shown schematically in Fig.~\ref{fig1}(b). In contrast to the one-dimensional case described in the previous section, the source now has a finite transverse dimension $\delta{y}$ along the $\hat{y}$ direction. The finite transverse dimension has a profound influence on the nature of the induced excitations as well as on the interactions of these structures with the background plasma. 
\begin{figure}[t]
\center
\includegraphics[width=1.0\textwidth]{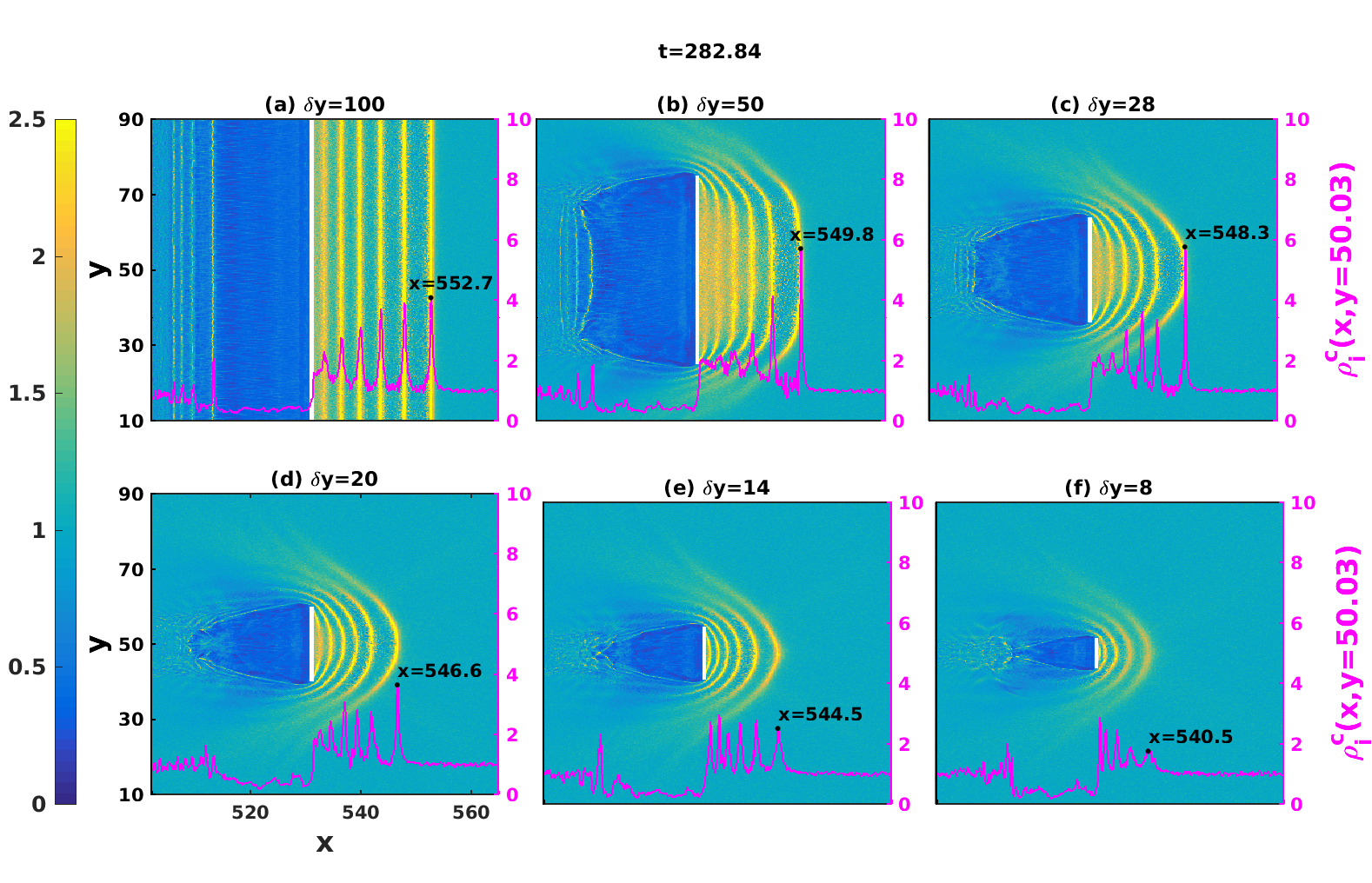}
\caption{The snapshots of the background ion charge density $\rho^{c}_i(x,y)$ for different ion sources of various transverse lengths $\delta{y}=100, 50, 28, 20, 14$ and $8$ at $t=282.84$. The vertical white line shows the location of the source.}
\label{fig5}
\end{figure} 
A quick comparative overview of the influence of the transverse length on the nature of the wave excitations is shown in Fig.~\ref{fig5} where snapshots of the background ion charge density $\rho^{c}_i(x,y)$ are presented at $t=282.84$  for simulations with $\delta{y}=100, 50, 28, 20, 14$ and $8$ respectively. All other parameters are kept the same for these runs. The vertical white line shows the location of the source. One of the first differences one notices is in the shapes of the precursors which now bend back at the tips to give rise to a crescent like structure. The bending becomes more pronounced as the transverse length of the source diminishes - a distinctly hydrodynamic phenomenon as the movement of the plasma flow around the source tries to attain a streamlined form. Note that this is different from a hydrodynamic bow wave that is normally formed around a ship's prow \cite{lamb1993hydrodynamics} or the plasma structure observed in PIC simulations, around an intense short laser pulse propagating in an underdense plasma \cite{esirkepov2008bow}. These bow waves remain attached to the moving source quite unlike our precursor structures which separate from the source and move away from it at a faster speed. We also notice  a slight decrease in the speed of the precursor with a reduction in the transverse size of the source. This can be discerned (as marked in Fig.~\ref{fig5}) from the distances travelled by the peaks of the leading precursors for the different cases within the same fixed period of time. The physical reason for this diminishment in speed can be attributed to the decrease in the charge strength of the source as the transverse length decreases. The decrease in the strength of the source is also responsible for the fewer number of precursors generated for the $\delta{y}=8$ case compared to the others and the decrease in the number of wakes. One also notices a reduction in the size of the ion depleted region - a geometric consequence of the decrease in the transverse source length. This ``bubble'' region (a terminology borrowed from the plasma wakefield acceleration literature) now displays both longitudinal and transverse inhomogeneity in density.

\begin{figure}[t]
\center
\includegraphics[width=1.0\textwidth]{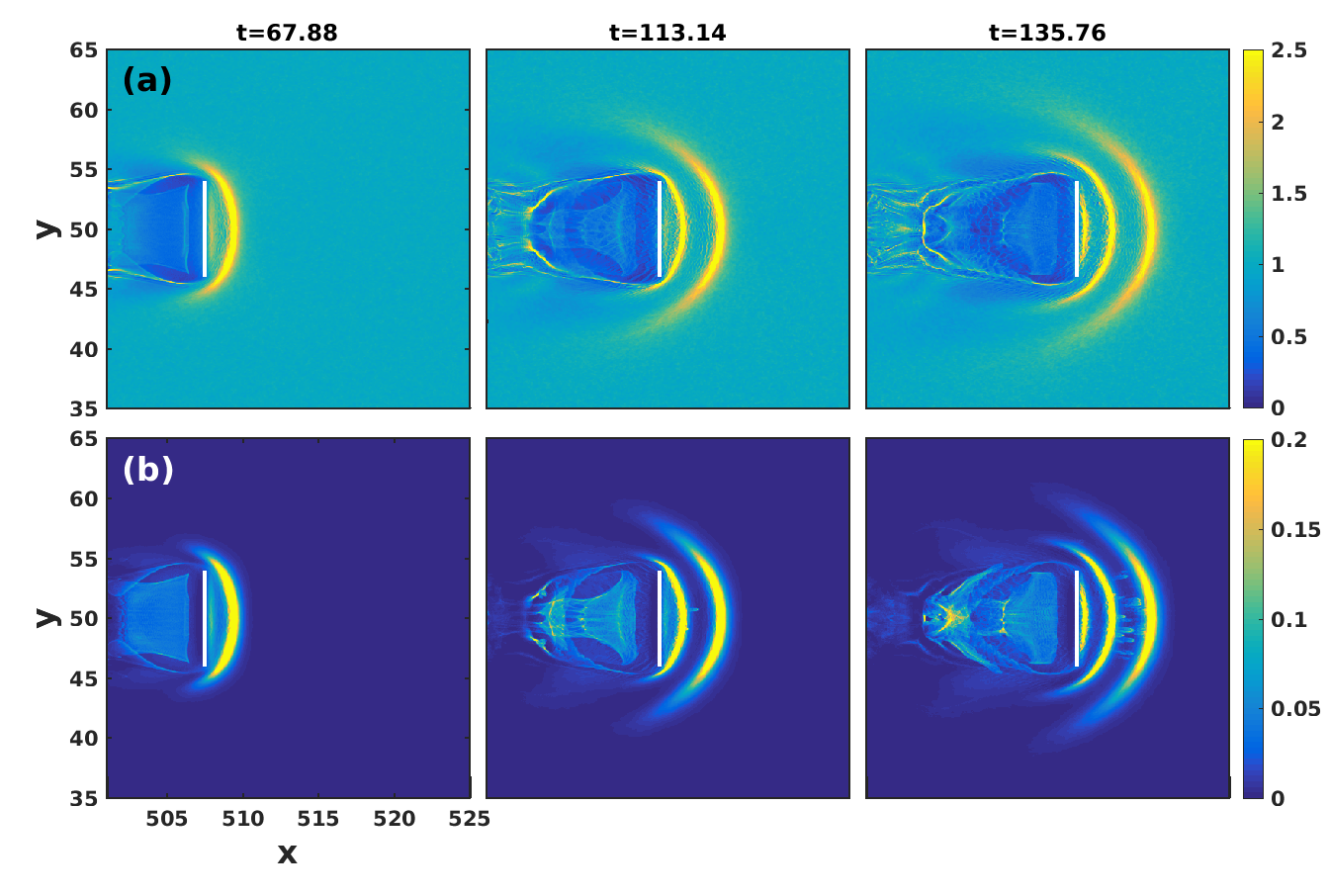}
\caption{The evolution of (a) the ion charge density $\rho^{c}_i(x,y)$ at various times while (b) displays the corresponding ion kinetic energy $K_i(x,y)$. Self-injection mechanism in the bubble regime where ions have been evacuated (dark blue) and regions where ions accumulate (light blue). The arrows show how ions are deflected outward and then accumulate at the back of the wakefield, where some of them are trapped and accelerated.}
\label{fig6}
\end{figure}

We will now take a closer look at the  $\delta{y}=8$ case in order to discuss in detail the dynamics of the background particles in the various regions and further delineate the nature of the wave structures. Figure~\ref{fig6}(a) shows a series of snapshots of the ion charge density $\rho^{c}_i(x,y)$ for this case, taken at various times while Fig.~\ref{fig6}(b) displays the corresponding snapshots of the ion kinetic energy $K_i(x,y)$. 
One observes a number of interesting features in Fig.~\ref{fig6} both in the bubble region and in the precursor region. As the beam propagates the bubble region starts filling up with particles that are self-injected from the background plasma. This is evident at $t=113.14$ and beyond, through the appearance of an increase in the ion population (shown in light blue in the plots of $\rho^{c}_i(x,y)$) in the evacuated bubble region that was originally in dark blue. In Fig.~\ref{fig6}(b) one also sees evidence of the energization of these self-injected ions inside the bubble region and a focusing effect leading to the formation of a beam like structure. The injected ions enter through the tail region of the bubble where they are swept in from the region ahead of the beam by the action of the  curving precursor waves. Such a self-injection is not seen in the 1d case since the ions ahead of the beam cannot go around it due to the infinite length of the beam. The physical phenomenon of self-injection and energization of the ions within the bubble is analogous to what happens to electrons in a wake-field acceleration scheme. Another interesting feature, distinctly visible in the rightmost plot of  Fig.~\ref{fig6}(b) is the trapping of particles between two precursors. These particles get energized by bouncing back and forth between the two wave fronts.
\begin{figure}[t]
\center
\includegraphics[width=1.0\textwidth]{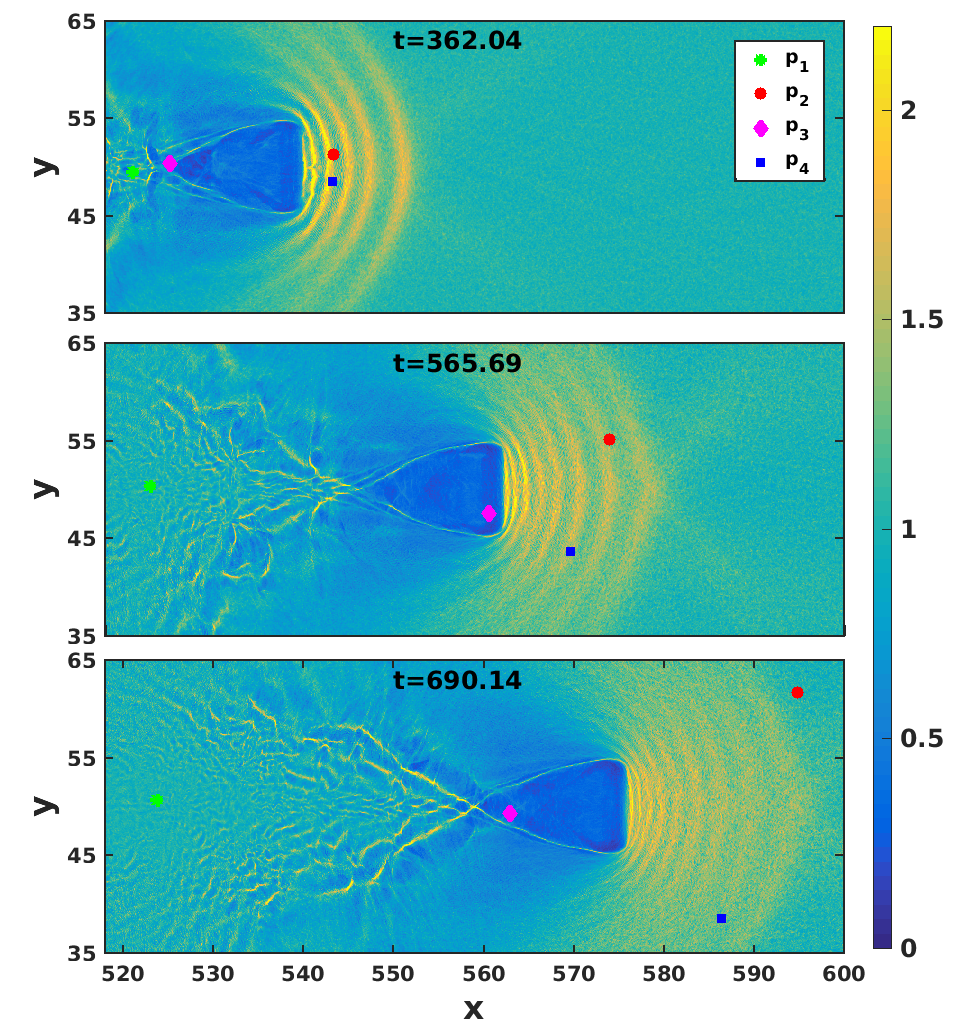}
\caption{Tagging and tracking. $p_1$, $p_2$, $p_3$ and $p_4$ are the four tagged particles. Initially $p_2$ and $p_4$ are the particles tagged between the two precursors and $p_1$ and $p_3$ are tagged at the back end of the bubble. Out of these four particles, one particle $p_3$ gets accelerated in the ion bubble/wake-field region and particle $p_4$ gets
energized and eventually gets untrapped.}
\label{fig7}
\end{figure} 
This is brought out more clearly in Fig.~\ref{fig7} where the progress of four tagged particles $p_1$, $p_2$, $p_3$ and $p_4$ are monitored. At $t=362.04$, these four particles are tagged at different location of the simulation domain: $p_2$ and $p_4$ are the particles tagged between the two precursors whereas $p_1$ and $p_3$ are tagged at the back end of the bubble. The time evolution of these particles shown in Fig.(\ref{fig7}) explains their overall dynamics. The particles $p_2$ and $p_4$ originally tagged between the two precursor structures get entrapped.  As time progresses, they bounce back and forth between those two wave fronts, get energized and  eventually in this process, they get untrapped. They are then pushed ahead of the precursors in the forward direction. These particles then travel faster than the precursors.  In certain cases groups of ions undergo such an acceleration and emerge in the form of thin beam-lets. Traces of the formation of such beam-lets were observed in the 1d case, as remarked earlier, but the acceleration is rather weak in that case. Furthermore, the particle $p3$ which is placed exactly at the back end of the bubble sees a huge electrostatic force acting in the forward direction. This force pulls this particle inside the bubble and accelerates it in the forward direction. The particle $p1$ which was placed slightly away from the back end of the bubble does not suffer any kind of acceleration. 

\begin{figure}[h]
\center
\includegraphics[width=1.0\textwidth]{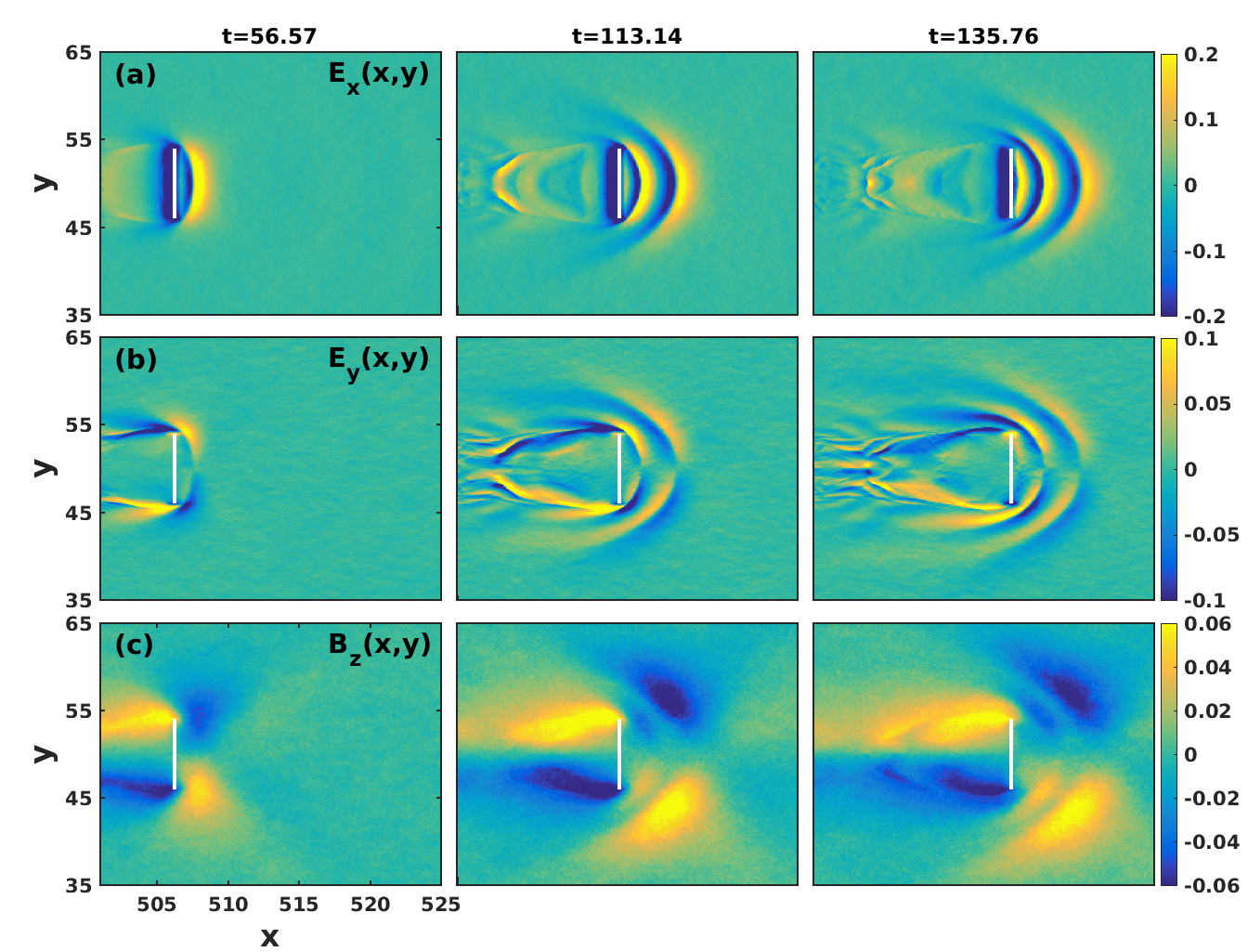}
\caption{The field components of the excited waves: (a) $E_x(x,y)$, (b) $E_y(x,y)$ and (c) $B_z(x,y)$.}
\label{fig8}
\end{figure}

 Finally we discuss one more marked difference between the 1d case and the present 2d simulation. This is in the nature of the wave excitations. Figure~\ref{fig8} shows the field components of the excited waves. 
 Unlike the 1d case we now have both $E_x$ and $E_y$ and also a finite magnetic field component $B_z$. In other words, the waves are no longer purely (or predominantly) electrostatic but have a non-negligible magnetic component. This is seen in all regions of the wave propagation -  precursors, bubble region and the wake region. The electromagnetic nature of the excitations is primarily due to the finite size of the source and in accord with past theoretical predictions \cite{yadavalli1965electromagnetic,bera2020effect, das2020boundary} and a recent experimental observation \cite{das2020boundary, chatterjee2017magnetic}.

\section{Discussion}
\label{disc}
In this work we have investigated the nature of linear and nonlinear collective excitations arising from the passage of an energetic ion bunch in a plasma. Furthermore, we have examined the interaction of these waves with the ambient plasma. These excitations include both trailing wakes and fore-wake structures. While trailing wakes excited behind a moving source have received a lot of attention in the past for particle acceleration applications, the fore-wake structures in plasmas have remained relatively unexplored. Our investigations are concerned with such scenarios where both trailing wake and fore-wake excitations can occur,  as distinct from the earlier works. Such structures are likely to arise whenever the source speed exceeds the phase speeds of some of the basic collective modes of the system.  In the present study we have restricted ourselves to cases where the beam speed exceeds the ion acoustic speed of the plasma and where the plasma is unmagnetized. We have then varied the transverse dimension of the beam to create different configurations of the source ranging from a nearly one dimensional line source to various two dimensional thin rectangular sources. We find that the shape and size of the beam have a profound influence on the nature of the precursor waves. Reducing the transverse dimension of the source causes a slowing down of the precursor and also bends it at the edges to create a crescent like structure. Since the charge density of the beam is kept constant, reducing the transverse length reduces the area of the source and hence the amount of charge on it. The reduction of the charge and hence the strength of the source leads to a reduction of the speed of the precursor and also to the number of precursors created in a given amount of time. The bending of the precursors is a geometric effect directly related to the shape of the source influencing the flow of the plasma around it. The finite transverse size of the source also introduces a modification in the field characteristics of the precursor and other source induced excitations in that they are no longer purely electrostatic but have a magnetic component to it. Such a modification arising from finite size effects had been predicted theoretically \cite{yadavalli1965electromagnetic,bera2020effect,das2020boundary} and observed experimentally \cite{chatterjee2017magnetic} in other contexts. This is the first time that they have been observed for precursor waves. \\

As regards the interaction of these waves with the ambient plasma we observe some interesting features some of which have a commonality with phenomena well known from wake-field particle acceleration studies. These include the formation of a bubble region and the trapping of ions and their subsequent acceleration inside the bubble. Our particle tagging studies demonstrate this very clearly. The formation of beam-lets provides a distinct signature of such an acceleration and focusing within the bubble region. The existence of the precursors, a novel feature of the present scenario, provides an additional mechanism for particle trapping and energization of ions. The basic mechanism is akin to Fermi acceleration due to repeated bouncing off these intense field structures. We also see evidence of beam-let formation in the fore-wake region arising due to particle trapping and acceleration in regions between precursors.

We now discuss some of the physical implications and potential applications of our results. Beam plasma interactions are a common feature of many natural phenomena and laboratory experiments. The particular circumstance of the beam traveling at a speed faster than the plasma sound speed (or the magneto-sonic speed in a magnetized plasma) is not difficult to realize. The supersonic component of the solar wind interacting with the earth's magnetosphere is a prime example in nature where such precursor excitations can take place. These would take the form of upstream travelling (towards the sun) nonlinear structures that can be detected from space-crafts. Space debris, inactive material objects ranging in size from microns to several centimeters that become highly charged and are traveling at orbital velocities, can excite such precursors. Both the precursors and the secondary signatures created by them in the ambient plasma could prove useful in tracing and predicting the trajectory of these objects. Energetic ion beams are also commonly used in various plasma heating experiments  {\it e.g.} in tokamaks where a neutral beam is injected across the magnetic field which then gets ionized to create an energetic ion beam. The ion beam delivers its energy directly to the plasma ions through a collisional process. Our results suggest a possible means of augmenting this heating through exploitation of the additional mechanism of wake-field/precursor  acceleration process \cite{magee2019direct} . Though our present simulations are restricted to unmagnetized plasmas they can be easily extended to magnetized plasmas where in the past we have demonstrated the existence of magneto-sonic precursors \cite{kumar2020precursor}. Such an investigation is presently in progress and will be reported separately.\\

\section{Acknowledgements}
\label{ack}
A.S. acknowledges AOARD for their research grant FA2386-18-1-4022 and is grateful to the Indian National Science Academy for the Honorary Scientist position. This  work is partially supported by  the U.S. Department of Energy, Office of Science, Office of Fusion Energy Sciences, under Contract No. DEAC05-00OR22725. The authors would like to acknowledge the OSIRIS Consortium, consisting of UCLA and IST (Lisbon, Portugal) for providing access to the OSIRIS4.0 framework which is the work supported by NSF ACI-1339893. The simulations for the work described in this paper were performed on Antya, an IPR Linux cluster.





\end{document}